Design of Porous Metal-Organic Frameworks for Adsorption Driven Thermal Batteries

Daiane Damasceno Borges[1], Guillaume Maurin[2], and Douglas S. Galvão[1]
[1]Applied Physics Department, University of Campinas - UNICAMP, Campinas-SP 13083-959, Campinas-SP, Brazil
[2]Institut Charles Gerhardt Montpellier UMR CNRS 5253, Université Montpellier 2, 34095 Montpellier cedex 05, France

ABSTRACT

Thermal batteries based on a reversible adsorption/desorption of a working fluid (water, methanol, ammonia) rather than the conventional vapor compression is a promising alternative to exploit waste thermal energy for heat reallocation. In this context, there is an increasing interest to find novel porous solids able to adsorb a high energy density of working fluid under low relative vapor pressure condition combined with an easy ability of regeneration (desorption) at low temperature, which are the major requirements for adsorption driven heat pumps and chillers. The porous crystalline hybrid materials named Metal–Organic Frameworks (MOF) represent a great source of inspiration for sorption based-applications owing to their tunable chemical and topological features associated with a large variability of pore sizes. Recently, we have designed a new MOF named MIL-160 (MIL stands for Materials of Institut Lavoisier), isostructural to CAU-10, built from the assembly of corner sharing aluminum chains octahedra $AlO_4(OH)_2$ with the 2,5-furandicarboxylic linker substituting the pristine organic linker, 1,4-benzenedicarboxylate. This ligand replacement strategy proved to enhance both the hydrophilicity of the MOF and its amount of water adsorbed at low $p/p_0$. This designed solid was synthesized and its chemical stability/adsorption performances verified. Here, we have extended this study by incorporating other polar heterocyclic linkers and a comparative computational study of the water adsorption performances of these novel structures has been performed. To that purpose, the cell and geometry optimizations of all hypothetical frameworks were first performed at the density functional theory level and their water adsorption isotherms were further predicted by using force-field based Grand-Canonical Monte Carlo simulations. This study reveals the ease tunable water affinity of MOF for the desired application.

INTRODUCTION

The increasing demand of heating and cooling buildings and vehicles for domestic and industry proposals have pushed scientific and engineer communities to develop new technology for replacing traditional heater-cooler devices. In this context, thermal battery based on the adsorption/desorption mechanism using clean working fluids such as water appear to be a promising alternative [1,2]. Although the physical principle of these batteries is already well-known and they have been commercialized since 19[th] century, their low efficiency and lack of environment-friendly adsorbate/adsorbent pairs makes this technology obsolete. Nowadays, the substantially increasing of efficiency is possible thanks to the emergence of new highly porous materials that can store large amounts of working fluid per volume unit. With the vast choice of adsorbent–adsorbate pairs, the selection of the most appropriate systems is nontrivial. The good pair candidate depends on the amount of heat that can be extracted from the evaporator per adsorption cycle, which is proportional to the amount of vapor adsorbed and the evaporation enthalpy of the selected working fluid. The selected working fluid should have a high enthalpy of evaporation while the capacity of the adsorbent should be maximized. Moreover, adsorption should take place at a typical relative working pressure of $p/p_0 = 0.17$ to 0.2 and the desorption hysteresis must be as small as possible or inexistent. The adsorbate could be water, methanol, ethanol or ammonia. The adsorbents like silica gel or zeolites (e.g. silica aluminophosphate SAPO-34) are quite hydrophilic, but exhibit a comparatively low water vapor uptake (e.g. silica gel approx. 0.2 g/g) [3].

Metal-organic framework (MOF) is a class of crystalline hybrid porous material that represents a great source of potential structures for sorption applications, because of their tunable chemical and topological features [4]. Most of these solids are superior to other microporous materials in terms of internal surface area and micropore volume. In the context of the application tackled in this contribution, some of these solids can be promising owing to their possible combination of large water adsorption uptake and excellent hydrothermal stability. Concerning this last requirement, Al-based MOFs have shown to be among the most stable hybrid materials in aqueous environment, besides while presenting low cost and low toxicity [5]. Recently, a few groups demonstrated that Al carboxylate based MOFs including CAU-10 (CAU stands for Christian-Albrechts-University) [6,7] and the Al fumarate A520, which is commercialized by BASF [8], show high water adsorption/desorption performance due to their high hydrophilicity and large water uptake. The hydrophilicity of CAU-10 can be improved during post-syntheses treatment by grafting functional groups such as -OH and -NH$_2$ on the organic linkers [9]. This method allows the enhancement of the material affinity for water, however this drastically drops the water uptake since the grafting leads to a decrease of the pore volume. An alternative strategy we propose is to replace the original organic linker, 2,4-benzenedicarboxylate in CAU-10 by a less bulky aromatic rings containing a polar heteroatom [10]. The corresponding Al carboxylate structures are thus constructed with the objective to not only enhance the hydrophilic character of the solid but also to maintain a highly accessible pore volume for the water molecules. In this short communication, we report preliminary results of an extended series of new MOFs materials derived from the parent CAU-10(Al) material. These structures are optimized by Density Functional Theory calculations and Grand Canonical Monte Carlo simulations are performed to predict their water adsorption performances.

**THEORY**

The structures that we have proposed were built up starting with the crystal structure of CAU-10 previously reported in the literature [6]; the benzene di-carboxylate ligand was substituted by six different di-carboxylate aromatic rings, *i.e.,* furan, pyrrole, thiophene, pyran, pyridine and pyrazine. Each structure was further geometry optimized using Density Functional Theory (DFT) *ab initio* methods. These calculations were performed using the Quickstep module of the CP2K code [11] within the generalized gradient approximation (GGA) with the Perdew-Burke-Ernzerhof (PBE) [12] functional. The van der Walls (dispersion function) corrections were considering with a cut-off radius of 10 Å. The cell optimization was carried out allowing the cell vector parameters *a,b* and *c* to change while keeping angles between the cell vectors as constant. The pressure tolerance was set equal 0.1bar. Once obtained these optimized structures, single point energy calculations and Mulliken [13] analysis (using the commercially available

DMol³ code [14]) were carried out to extract the partial charge carried by each atom of the MOF framework.

These optimized structures were further used to model the water adsorption properties by GCMC simulations using the Complex Adsorption and Diffusion Simulation Suite (CADSS) code [15]. The simulation boxes were composed of 2 × 2 × 4 unit cells, with a rigid framework interacting with water molecules through 12-6 Lennard–Jones (LJ) potential and Coulombic forces. The TIP4P/2005 model [16] was used to represent the water molecule while the LJ interatomic potential parameters for describing each atom of the framework were taken from the Universal force field (UFF) [17] parameters with two modifications: first, the interaction with the inorganic part was described without considering the LJ interactions with Al atoms, as defined and justified in previous studies for MIL-53(Al) [18]; second, the LJ contributions from the H atoms from hydroxyl groups and organic linkers were not considered. This approach was adopted in our previous work and it shows a very good agreement with experimental data [10].

## DISCUSSION

The Al-based MOF called CAU-10 is composed of inorganic aluminum octahedral chains linked via carboxylate groups of the ligand 2,4-benzenedicarboxylate (BDC) (see Figure 1). Each $AlO_6$ octahedron is surrounded by four ligands and two hydroxyl groups. The two –OH ions are in *cis-* position and bridge the Al centers to create the chains. These helical chains, previously observed in a Sc-based MOF [19], run along the *c*-axis to form 1D porous network of square-shaped channels of ≈5.6 Å in diameter.

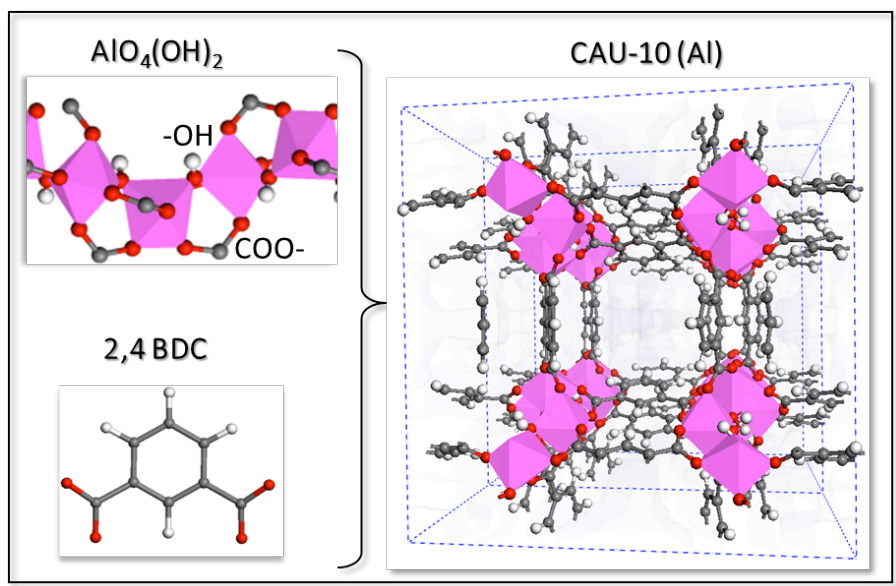

**Figure 1.** Illustration of a CAU-10(Al) framework: left-up side shows the connected Al helical chains cis-connected and surrounded by four carboxylate groups from four ligands; left-down shows the organic ligand 2,4-benzenedicarboxylate (BDC); right side shows the square-shaped channels delimited via the ligands connecting four chains. Al polyhedra, carbon, oxygen, and hydrogen atoms are in pink, gray, red and white colors, respectively.

**Table 1.** DFT optimized lattice parameters for the MOFs obtained using the CP2K software package. Each structure differs only by its organic linker.

| | | a(Å) | b(Å) | c(Å) | α |
|---|---|---|---|---|---|
| CAU-10 | 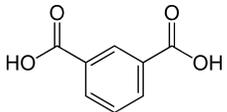 | 21.657 | 21.657 | 10.795 | 90 |
| CAU-10 Furan | 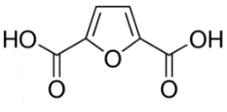 | 21.230 | 21.231 | 10.642 | 90 |
| CAU-10 Pyrrole | 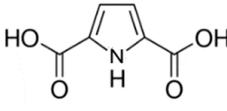 | 21.353 | 21.354 | 10.629 | 90 |
| CAU-10 Thiophene | 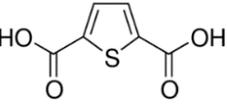 | 22.032 | 22.032 | 10.618 | 90 |
| CAU-10 Pyran | 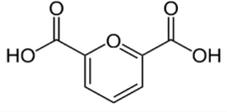 | 21.307 | 21.305 | 10.754 | 90 |
| CAU-10 Pyridine | 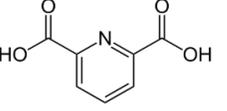 | 21.352 | 21.353 | 10.756 | 90 |
| CAU-10-2,4 Pyrazine | 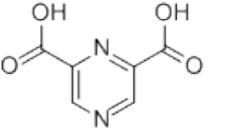 | 21.335 | 21.337 | 10.847 | 90 |

**Table 2.** Accessible solvent surface area computed using Material Studio [20]. Pore volume and mass density values are also presented.

| | CAU-10 | CAU-10 Furan | CAU-10 Pyrrole | CAU-10 Thiophene | CAU-10 Pyran | CAU-10 Pyridine | CAU-10 Pyrazine |
|---|---|---|---|---|---|---|---|
| Surface Area (Å²) | 713.10 | 776.15 | 779.81 | 860.71 | 672.71 | 732.00 | 809.63 |
| Pore Volume (cm³/g) | 0.43 | 0.45 | 0.45 | 0.47 | 0.39 | 0.43 | 0.42 |
| Mass density (g/cm³) | 1.0971 | 1.0971 | 1.0804 | 1.1038 | 1.1488 | 1.1328 | 1.1304 |

For the new structures, the 2,4 BDC were computationally replaced by six different heterocyclic linkers. In Tables 1 and 2 we present the DFT optimized lattice parameters of the conventional unit cell and relevant structural data for each structure considered here. We can infer from these tables that the framework volumes are very similar for all structures and the mass densities remain around 1.1g/cm3 (see Table 2). The *a* and *b* parameters varies less than 3% in comparison to pristine CAU-10, while the *c* parameter vary less than 2%. Table 2 shows the calculated pore volume available for Helium adsorption [21] using CAADS code and the accessible solvent surface area computed using the module *"Atom Volumes & Surfaces"* implemented in Material Studio software [20]. The CAU-10Thiophene presents the highest surface area of 860.71 Å² and the highest pore volume of 0.47 cm$^3$/g, while the CAU-10Pyran presents the smallest values of surface area, equals to 672.71 Å², and pore volume equals to 0.39 cm$^3$/g.

The water sorption behavior was further investigated for each structure. GCMC simulations were carried out by varying the water vapor reservoir pressure to predict the amount of water molecules adsorbed in each solid at ambient temperature T=300K. Figure 2 shows the water adsorption isotherms and the corresponding adsorption enthalpies for each case studied. Notice that the s-shape isotherm is observed for all case except for the CAU-10Pyran, which has a very strong hydrophilicity. As the hydrophilicity of water adsorbent increases, the sorption rate could increase but water desorption could retard leading to higher desorption temperature, which leads CAU-10Pyran to be unappropriated for thermal battery application. In contrast, the CAU-10Pyrrole is the most hydrophobic, with the adsorption starting point at relatively high pressure (> 0.2 p/p$_0$), which is also not appropriate for battery applications. The best candidates are the -Furan, -Pyridine and -Pyrazine that have considerably improved the water affinity and adsorption capacity compared to CAU-10 (0.37 g/g dry MOF *vs* 0.32 g/g). The -Thiophene also present an excellent candidate. Although it has similar water affinity compared to CAU-10, the adsorption capacity is high (0.42 g/g). Further validation of these results, as well as the molecular understanding of adsorbed-adsorbent interactions will be provided in our future publication.

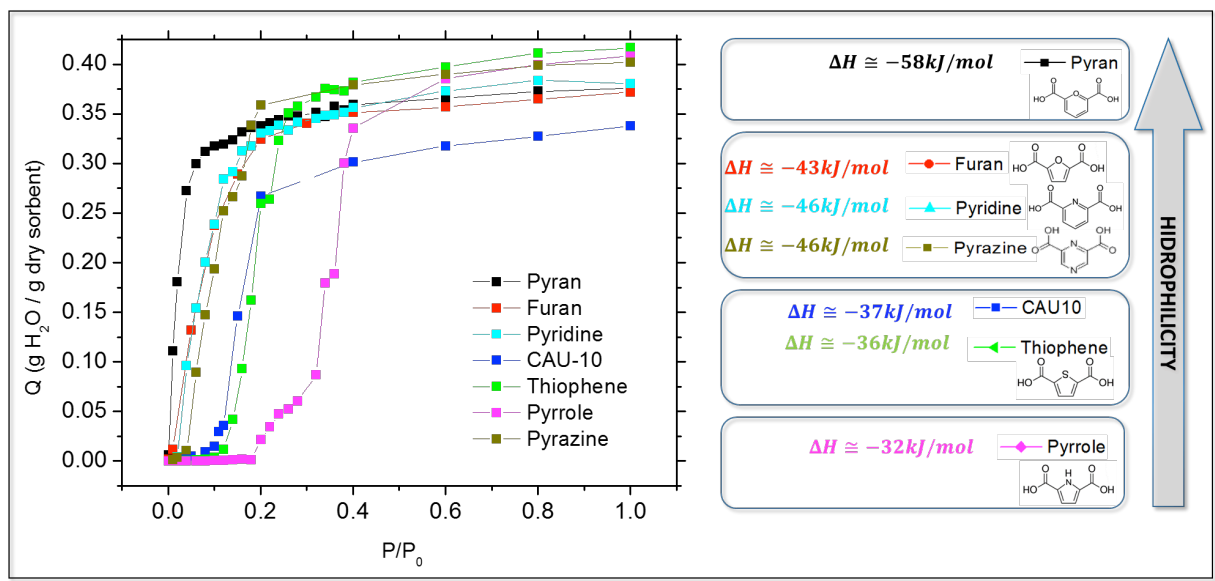

**Figure 2.** Water adsorption isotherm curves (left) and the calculated adsorption enthalpies (right) at temperature 300K.

# CONCLUSIONS

Computational studies were performed to design a series of CAU-10 inspired new MOF material for thermal battery applications. Water sorption isotherms and enthalpies were predicted. We have shown that although the topology and the pore volumes are very similar in all cases, the type of organic linker strongly affect the water affinity and adsorption capacity of the material. Our systematic study has proofed the tunable character of MOFs to improve sorption properties for industry application.

# ACKNOWLEDGMENTS

This work was supported in part by the Brazilian Agencies CAPES, CNPq and FAPESP. The authors also thank the Center for Computational Engineering and Sciences at Unicamp for financial support through the FAPESP/CEPID Grant # 2013/08293-7, # 2015/14703-9.